\documentclass[]{aastex631}
\usepackage{appendix}
\usepackage{datatool,booktabs}
\usepackage{subfigure} %

\usepackage{amsmath}

\def\aap{Astron.\ Astrophys.\ }

\def\apjl{Astrophys.\ J.\ Lett.\ }

\def\apjs{Astrophys.\ J.\ Supp.\ }

\newcommand{\zcr}[1]{{\color{black} {#1}}}

\begin{document}

\title {Forecasting  of the time-dependent fluxes of antiprotons in the AMS-02 era   }

\author{Cheng-Rui Zhu}
\email{zhucr@anhu.edu.cn}
 \affiliation{ Department of Physics, Anhui Normal University, Wuhu, Anhui, 241000, China}
 \affiliation{Key Laboratory of Dark Matter and Space Astronomy, Purple Mountain Observatory, Chinese Academy of Sciences, Nanjing 210023, China }
\author{Kai-Kai Duan}
\email{duankk@pmo.ac.cn}
\affiliation{Key Laboratory of Dark Matter and Space Astronomy, Purple Mountain Observatory, Chinese Academy of Sciences, Nanjing 210023, China }


\date{\today}

\begin{abstract}

The spectra of galactic cosmic rays (GCRs) contain crucial information about their origin and propagation through the interstellar medium. When GCRs reach Earth, they are significantly influenced by the solar wind and the heliospheric magnetic field, a phenomenon known as solar modulation. This effect introduces time-dependent variations in GCR fluxes. The AMS-02 experiment has released time-dependent flux data for protons, electrons, and positrons, revealing clear correlations with solar modulation. Studies suggest that cosmic rays with the same charge, such as protons and helium nuclei, exhibit similar/same solar modulation parameters. In this work, we derive the LIS for protons and positrons under the assumption of a common solar modulation potential, using data from Voyager 1 and a 7-year average from AMS-02. Similarly, the LIS for antiprotons and electrons is derived by assuming they are governed by a separate solar modulation potential. We demonstrate that the time-dependent fluxes of positrons and protons can be accurately modeled using the same set of solar modulation parameters within a modified force-field approximation framework. Based on this, we predict the time-dependent fluxes of antiprotons using the corresponding electron flux data.


\end{abstract}



\section{Introduction}

Galactic cosmic rays (GCRs) are charged, energetic particles originating from cosmic accelerators, such as supernova remnants, and propagate throughout the Milky Way~\citep{1998ApJ...493..694M}. Upon entering the heliosphere, they are modulated by the outward-moving magnetized solar wind plasma~\citep{Potgieter2013}. It is important to study the solar modulation as it is crucial for understanding the nature of GCRs, including their origins~\citep{2013A&ARv..21...70B} and propagation mechanisms~\citep{2007ARNPS..57..285S} within the galaxy. Additionally, solar modulation plays a critical role in the search for dark matter, as it affects low-energy antiproton and antideuteron fluxes~\citep{2012CRPhy..13..740L,Yuan:2014pka,2017PhRvL.118s1101C,2022PhRvL.129w1101Z}.

The propagation of GCRs within the heliosphere is described by Parker's transport equation (TPE)~\citep{1965P&SS...13....9P}. This equation can be solved using various methods, including numerical and analytical approaches. Usually, the force field approximation (FFA)~\citep{1967ApJ...149L.115G,1968ApJ...154.1011G} is used to solve the equation as it is simple and enough to explain the observations. 
The recent experimental results have achieved significant breakthroughs that are instrumental in understanding the solar modulation effect as well as the GCRs physics, particularly those from Voyager, AMS-02, PAMELA and DAMPE~\citep{2013Sci...341..150S,2011Sci...332...69A,2017PhRvL.119y1101A,cite-key}. To date, only Voyager-1 and Voyager-2 have crossed the heliospheric boundary~\citep{2013Sci...341..150S,voyager-2}, providing direct measurements of the LIS in the energy range of a few to hundreds of MeV/nucleon. The AMS-02 collaboration has further contributed by publishing high-precision cosmic ray spectra~\citep{AGUILAR20211} and the evolution of some cosmic ray fluxes over time~\citep{PhysRevLett.121.051101,PhysRevLett.121.051102}, providing new opportunities to study the solar modulation for cosmic rays (CRs).
The FFA cannot explain the fluxes itself as well as the changing over a long period of time with homogeneous parameters~\citep{2021ApJ...921..109S,PhysRevD.109.083009}. Several methods have been proposed to modify the FFA to account for the cosmic rays fluxes~\citep{2016ApJ...829....8C,2017PhRvD..95h3007Y,Zhu:2020koq,2021ApJ...921..109S,PhysRevD.106.063021,PhysRevD.109.083009}.

Recently, AMS-02 released time-dependent flux data for protons (positrons, and electrons) spanning from 2011 to 2020 (to 2022) \citep{AMS:2021qln,PhysRevLett.131.151002,PhysRevLett.130.161001}, 
offering new avenues to study solar modulation. In this study, we utilize two modified force-field approximation models to investigate the solar modulation of protons, positrons, and electrons. Both models successfully fit the daily and monthly fluxes of protons and positrons using the same set of parameters. Building on this, we predict the daily fluxes of antiprotons based on the corresponding electron fluxes data.

\section{Methodology}
\subsection{Solar Modulation}

The existence of heliospheric magnetic field carried by solar winds causes the modulation of GCRs as they enter the heliosphere, resulting in suppressed fluxes of CRs. This phenomenon is known as solar modulation, and its effects are more pronounced at particle energies below 30 GeV/n~\citep{Potgieter2013}. The basic transport equation was first derived by Parker~\citep{1965P&SS...13....9P}. 3D time-dependent self-consistent modelling is a full solution to the CRs transport problem in the heliosphere, however it is a complicated task and requires a very large amount of computation. The FFA is a simplified method to solve this problem. Usually, the FFA requires the quasi-steady changes, spherical symmetry, etc, which are apparently invalid for short time scales. In fact, they are not fully valid even for the regular condition ~\citep{2003JA010098}. However, the force-field formalism was found to provide a very useful and comfortable mathematical parametrization of the GCR spectrum even during a major Forbush decrease, irrespective of the (in)validity of physical assumptions behind the force-field model. So one can still benefit from the simple parametrization offered by the FFA for practical uses other than studying the physics of solar modulation, even if the FFA is not a physically motivated solution to the solar modulation problem at daily time scales~\citep{USOSKIN20152940}.

In this model, the top-of-atmosphere (TOA) flux is related to the local interstellar spectrum (LIS) flux as follows:  
\begin{equation}\label{force_filed}
J^{\rm TOA}(E)=J^{\rm LIS}(E+\Phi)\times\frac{E(E+2m_p)}
{(E+\Phi)(E+\Phi+2m_p)}, 
\end{equation}
where $E$ is the kinetic energy per nucleon, $\Phi=\phi\cdot Z/A$ with 
$\phi$ representing the solar modulation potential,
$Z$ and $A$ are the atomic number and mass number of the cosmic ray particle, respectively,
$m_p=0.938$ GeV is the 
proton mass, and $J$ denotes the differential flux of GCRs. The sole parameter in the FFA is the modulation potential $\phi$.


\subsection{Modified Force-Field Approximation}

The force-field model assumes a quasi-steady-state solution to Parker's transport equation. However, observational GCR fluxes exhibit 11-year variations linked to solar activity. To account for this, a time-series of $\phi$ at different epochs is used to describe the data. Since a single parameter cannot adequately fit the monthly cosmic ray fluxes, a rigidity-dependent solar modulation potential is required~\citep{SIRUK20241978}.  

In this study, we employ two modified force-field approximation  models. First, we use Zhu's model from~\cite{zhu2024}, an extension of the standard force-field approximation. The solar modulation potential in this model is defined as:
\begin{equation}\label{Zhu}
\phi (R)_{Zhu} = \phi_l +\left (\frac{\phi_h-\phi_l}{1+e^{(-R+R_b)}} \right ),
\end{equation}
where $\phi_l$ and $\phi_h$ are the modulation potentials for low and high energies, respectively, $R$ is the rigidity, $R_b$ is the break rigidity and $e$ is the natural constant. 
A sigmoid function is employed to ensure a smooth transition between $\phi_l$ and $\phi_h$, which is necessary to model the gradual change in modulation potential across different energy ranges. The parameters $\phi_l$, $\phi_h$, and $R_b$ are free and determined through fitting.

Second, we adopt Long's model from~\cite{Kuhlen_2019,PhysRevD.109.083009}, which introduces a power-law dependence on rigidity to account for variations in solar modulation. The solar modulation potential in this model is expressed as:  
\begin{equation}\label{Long}
\phi (R)_{Long} = \phi_0 + \phi_1 ln(R/R_0), 
\end{equation}
with 
\begin{equation}\label{force_filed3}
J^{\rm TOA}(E)=J^{\rm LISA}(E+\Phi) \times \frac{E(E+2m_p)}{(E+\Phi)(E+\Phi+2m_p)}exp(-g\frac{10R^2}{1+10R^2}\phi)).
\end{equation}
where $\phi_0$, $\phi_1$, and $g$ are free parameters to be fitted. The parameter $g$ controls the rigidity dependence, allowing the model to capture the energy-dependent modulation effects more accurately.

\subsection{Markov Chain Monte Carlo (MCMC)}

We fit the two  solar modulation $\phi(R)$ with three free parameters. The $\chi^2$ statistics is defined as
\begin{eqnarray}
\chi^2=\sum_{i=1}^{m}\frac{{\left[J(E_i;\phi(R))-
J_i(E_i)\right]}^2}{{\sigma_i}^2},
\end{eqnarray}
where $J(E_i; \phi(R))$ is the expected modulated flux, $J_i(E_i)$ is the measured flux, and $\sigma_i$ is the measurement error for the $i$-th data bin, with $E_i$ representing the geometric mean of the bin edges. The errors $\sigma_i$ include both statistical and systematic uncertainties, ensuring a robust fit.

To minimize the $\chi^2$ function, we employ the MCMC algorithm within a Bayesian framework. The posterior probability of the model parameters $\theta$ is given by:  
\begin{eqnarray}
p(\boldsymbol{\theta}|{\rm data}) \propto {\mathcal L}({\rm data}|\boldsymbol{\theta})
p(\boldsymbol{\theta}),
\end{eqnarray}
where ${\mathcal L}({\rm data}|\boldsymbol{\theta})$ is the likelihood function and 
$p(\boldsymbol{\theta})$ is the prior probability, chosen to reflect physical constraints on the parameters.

The MCMC driver, adapted from {\tt CosmoMC}~\citep{2002PhRvD..66j3511L,Liu_2012}, uses the Metropolis-Hastings algorithm. Starting from a random point in the parameter space, the algorithm proposes new points based on the covariance of the parameters, which is estimated from preliminary fits. The acceptance probability for a new point is defined as:  

\begin{eqnarray}
P_{\text{accept}} = \min\left(1, \frac{P(\theta_{\text{new}} | \text{data})}{P(\theta_{\text{old}} | \text{data})}\right). 
\end{eqnarray}  
If accepted, the process repeats from the new point; otherwise, it reverts to the previous point. This iterative process ensures thorough exploration of the parameter space, yielding robust estimates of the modulation parameters.
For more details about the MCMC one can refer to~\citep{MCMC}.

\subsection{LIS of Protons, Electrons, Positrons, and Antiprotons}

Typically, power-law or broken power-law functions are used to fit GCR data~\citep{2014A&A...566A.142Y}. However, if the observational data span a sufficiently wide energy range, a non-parametric method such as spline interpolation can be employed~\citep{2016A&A...591A..94G,2017AdSpR..60..833G,Zhu:2018jbk}. Spline interpolation constructs a smooth function passing through a series of points using piecewise polynomial functions. Here, we use cubic spline interpolation, with the highest polynomial order set to three, to ensure a smooth and accurate representation of the LIS~\cite{2016A&A...591A..94G,zhu2024}.

\begin{figure*}
    \centering
    \includegraphics[scale=0.7]{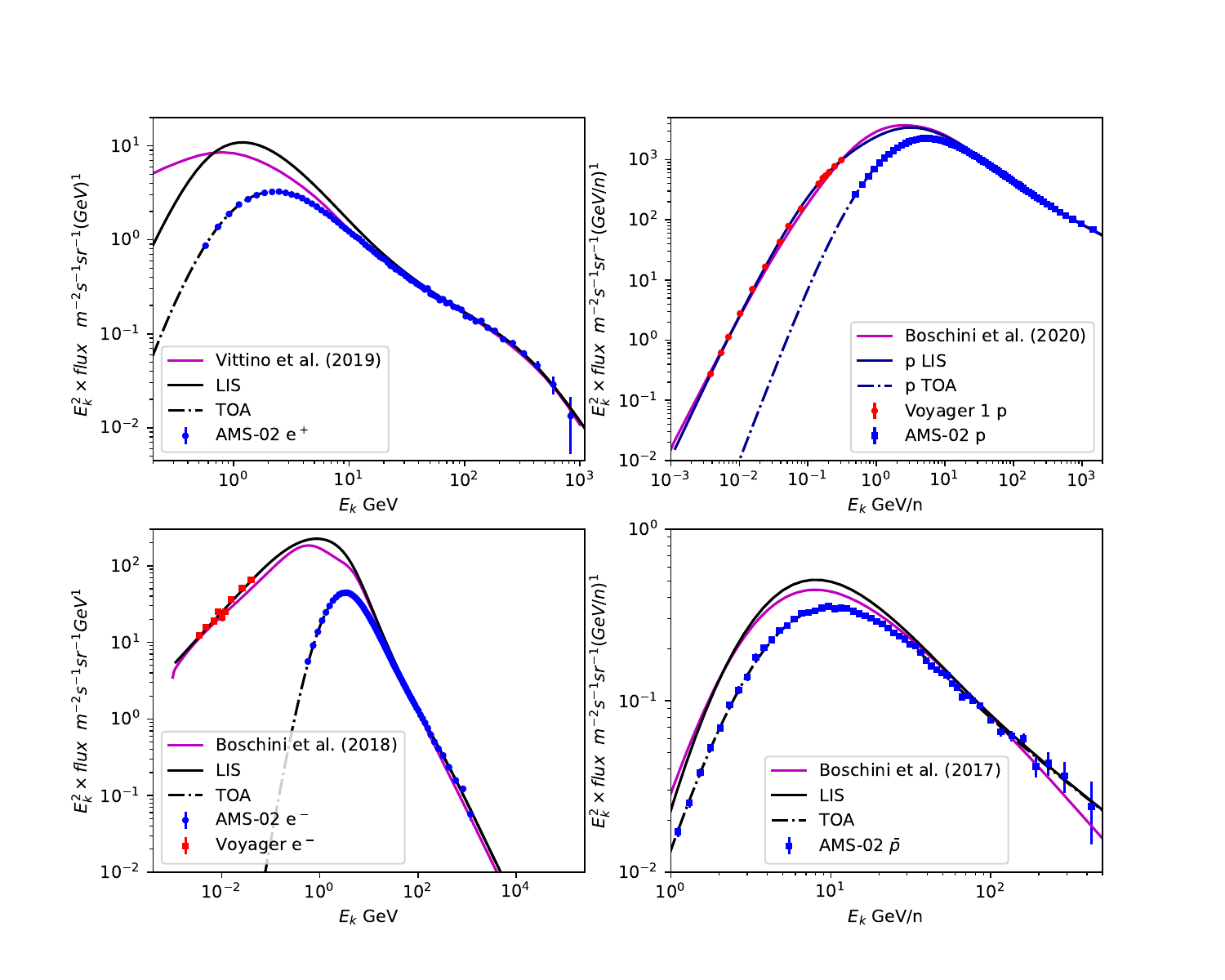}
    \caption{ LIS of positron (top left), proton (top right), electron (bottom left) and antiproton (bottom right). The blue points are AMS-02 data from~\citep{AGUILAR20211}, and the red points are Voyager data from~\cite{2016ApJ...831...18C}. The lines stand for the LIS while the dash lines donate TOA. \zcr{The black lines correspond to the LISs used in this work while the magenta curves represent reference LIS  from literature sources for comparison \citep{Boschini_2017,Boschini_2018,PhysRevD.100.043007,Boschini_2020}.}}
    \label{fig:LIS}
\end{figure*}

\zcr{Based on Voyager data for electrons and protons, we can successfully derive their LISs and corresponding solar modulation potentials using the FFA by assuming that Voyager data contain only protons or electrons, with no other components. Unfortunately, there is no Voyager data for positrons and antiprotons. So, in order to obtain the LISs for positrons and antiprotons, we assume that positrons share the same solar modulation potential as protons and that antiprotons share the same solar modulation potential as electrons, as they have the same charge sign. Since we have obtained the proton LIS in \cite{zhu2024,zhu2024b}, we use the same proton LIS here. We assume that positrons share the same solar modulation potential as proton in \cite{zhu2024} , which is 0.477 GV. Then, we fit the positron LIS while keeping the solar modulation potential unchanged. For the antiproton LIS, we fit it together with the electron LIS and make them share the same solar modulation potential during fitting. Once the LISs are established, they are held constant throughout the analysis. The fitting results are illustrated in Figure~\ref{fig:LIS} and we compare our best-fitting LIS with previous works with somehow different methods and assumptions. Our LISs are very closed to the results from Galprop-HelMod join effort \citep{Boschini_2017,Boschini_2018,Boschini_2020} for proton, electron and antiproton. The positron LIS used in this work is also close to \cite{PhysRevD.100.043007} at energy above 2 GeV.}

\section{results }

\subsection{ Fitting Time-Dependent Fluxes of Protons and Positrons }

\begin{figure*}
    \centering
    \includegraphics[scale=0.8]{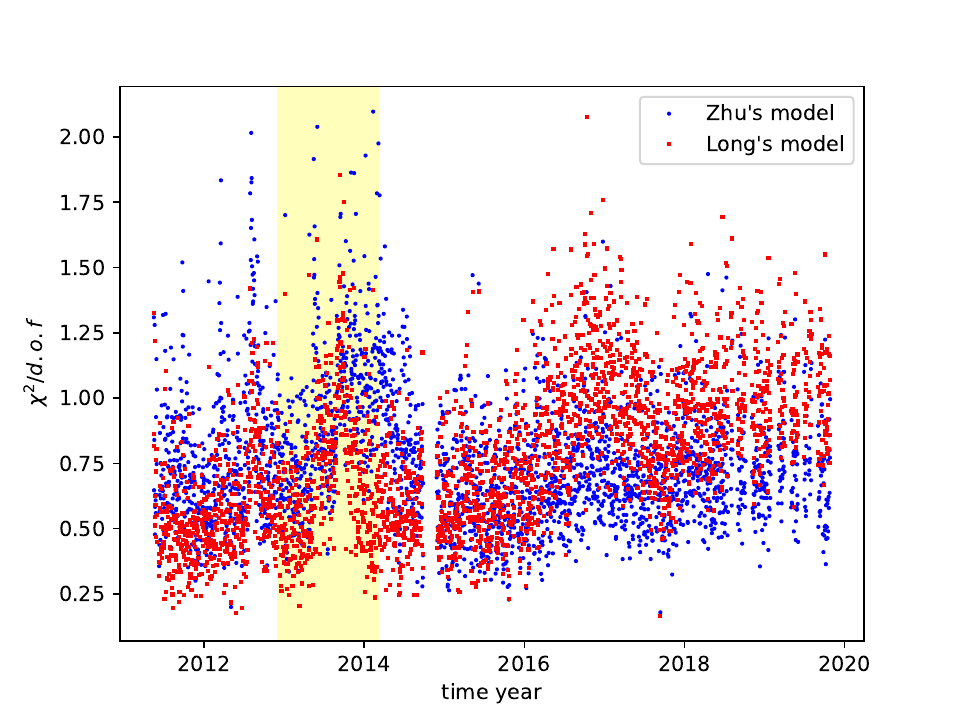}
    \caption{$\chi^2/d.of.$ values for Zhu's and Long's model using daily proton and positron fluxes. The yellow shaded band stands for the  heliospheric magnetic field reversal period within which the polarity is uncertain~\citep{2015ApJ...798..114S}.}
    \label{fig:chi2}
\end{figure*}

Using the derived LIS, we simultaneously fit the daily fluxes of protons and positrons with the solar modulation model. Since the standard FFA struggles to fit time-dependent fluxes~\citep{PhysRevD.109.083009,zhu2024}, we present the $\chi^2/\text{d.o.f}$ values for Zhu's and Long's models in Figure~\ref{fig:chi2}. Zhu's model yields $\chi^2/\text{d.o.f}$ values ranging from 0.179 to 2.097, with a mean of 0.756. Long's model produces values ranging from 0.173 to 2.130, with a mean of 0.786. Notably, we constrain the parameter $g$ in Long's model to the range $[-0.2, 0.2]$. Relaxing this constraint improves the fit, yielding a mean $\chi^2/\text{d.o.f}$ of 0.634, but at the cost of increased degeneracy in the results, as the parameters become less well-constrained.

\begin{figure*}
    \centering
    \includegraphics[scale=0.8]{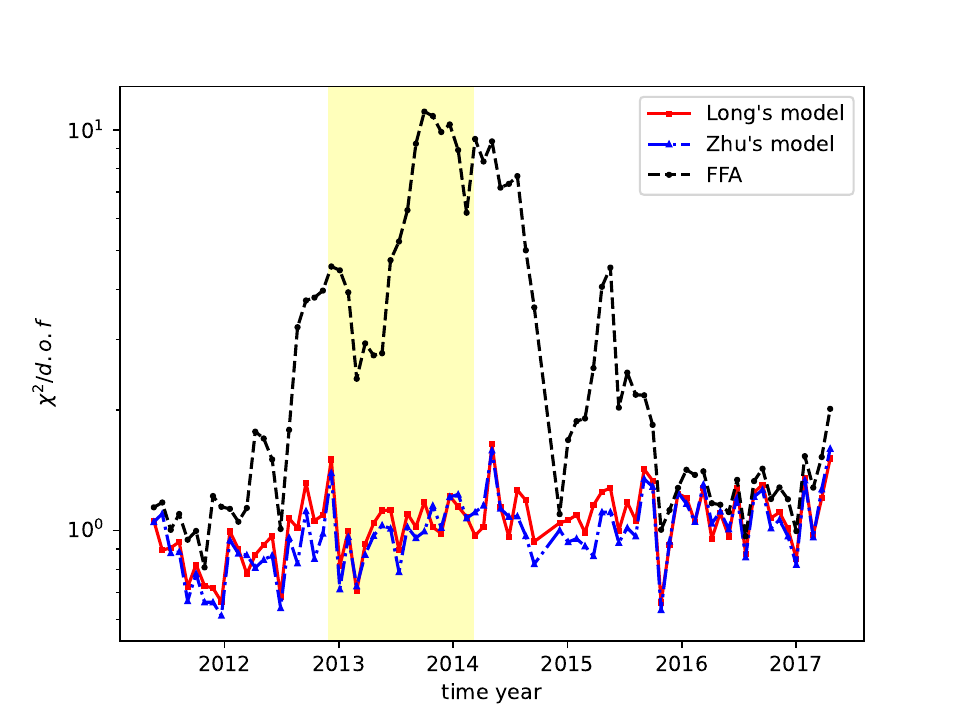}
    \caption{$\chi^2/d.of.$ values for Zhu's (blue) and Long's (red) model compared to the FFA (black) using monthly proton and positron fluxes. The yellow shaded band stands for the  heliospheric magnetic field reversal period within which the polarity is uncertain\citep{2015ApJ...798..114S}.}
    \label{fig:chi2_mn}
\end{figure*}

Since the daily positron fluxes consist of only 12 data bins with uncertainties of approximately 10\%, while the daily proton fluxes have more data bins with smaller uncertainties, the $\chi^2$ values and solar modulation parameters are predominantly influenced by the proton data rather than the positron data. This is also  why Helium fluxes are excluded from this analysis. To address this, we recalculated the results using monthly fluxes of protons and positrons, which have comparable data bins and uncertainties. The $\chi^2$/d.o.f values for Zhu's (blue) and Long's (red) models are presented in Figure~\ref{fig:chi2_mn}, alongside the $\chi^2$/d.o.f values for the FFA model for comparison. Both Zhu's and Long's models yield similar fitting results, with mean $\chi^2$/d.o.f values of 1.006 for Zhu's model and 1.053 for Long's model.

The $\chi^2/\text{d.o.f}$ values indicate that protons and positrons share the same solar modulation parameters within the AMS-02 rigidity range (approximately 1–100 GV), as they undergo similar diffusion processes in the heliosphere due to their identical charge sign. This similarity arises because particles with the same charge experience comparable interactions with the heliospheric magnetic field, leading to analogous modulation effects.

\begin{figure*}
    \centering
    \includegraphics[scale=0.6]{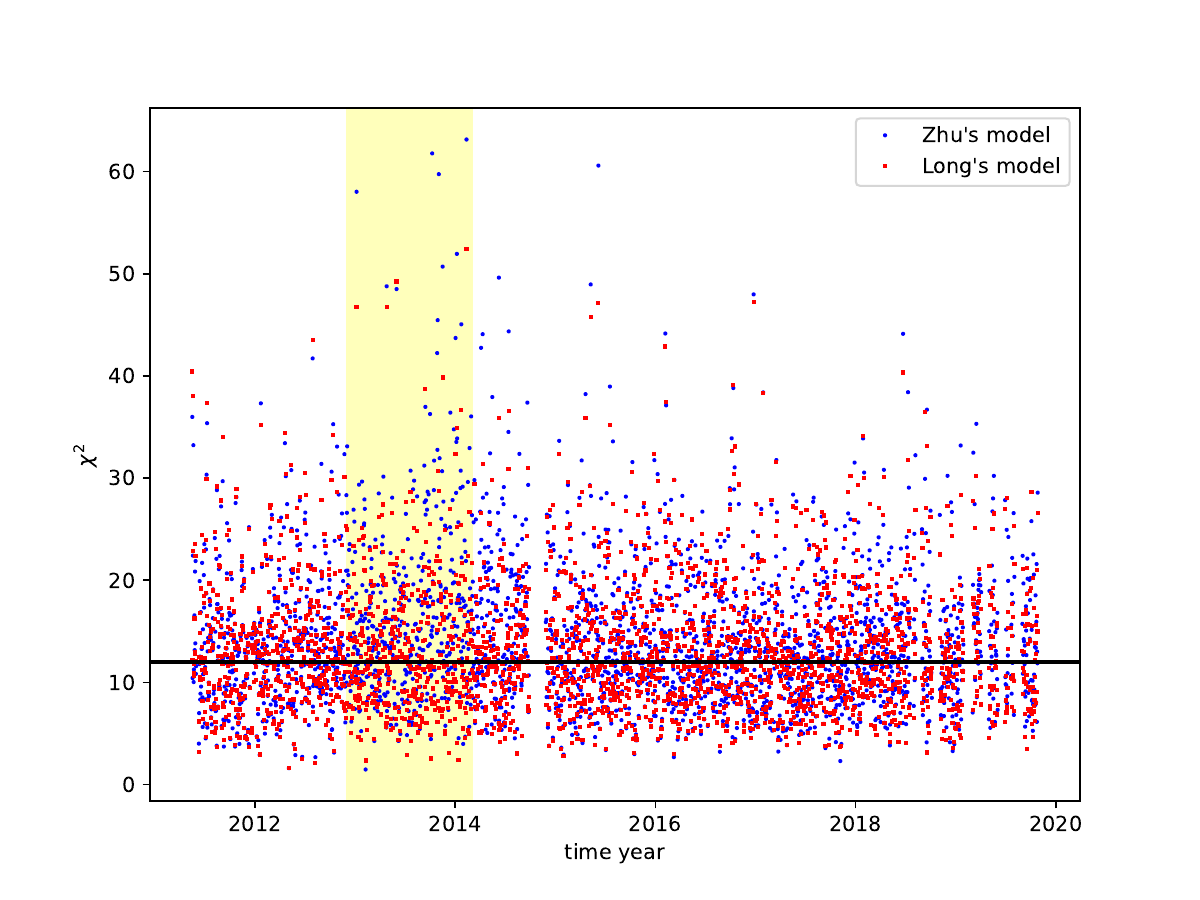}
    \caption{$\chi^2/d.of.$ values for Zhu (blue) and Long's (red) model with daily positron fluxes using the solar modulation parameters from~\cite{zhu2024b}. The black line denotes the data bin number for daily positrons fluxes, which is 12. The yellow shaded band stands for the  heliospheric magnetic field reversal period within which the polarity is uncertain~\citep{2015ApJ...798..114S}.}
    \label{fig:chi2_posi}
\end{figure*}

\begin{figure*}
    \centering
    \includegraphics[scale=0.7]{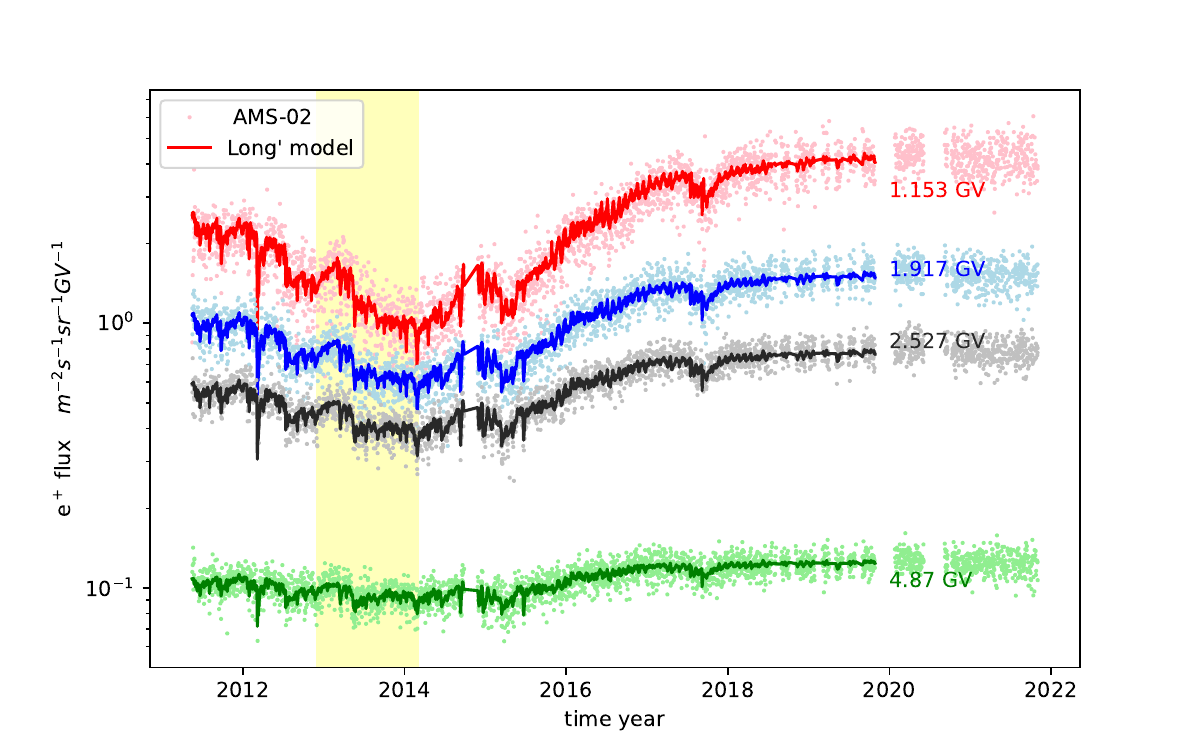}
    \caption{Long's model prediction (lines) of positron fluxes comparing to the AMS-02 data (points data) at Rigidities = 1.153 GV, 1.917 GV, 2.527 GV and 4.87 GV (from top to bottom) .  The yellow shaded band stands for the  heliospheric magnetic field reversal period within which the polarity is uncertain~\citep{2015ApJ...798..114S}.}
    \label{fig:posi_f}
\end{figure*}

\begin{figure*}
    \centering
    \includegraphics[scale=0.7]{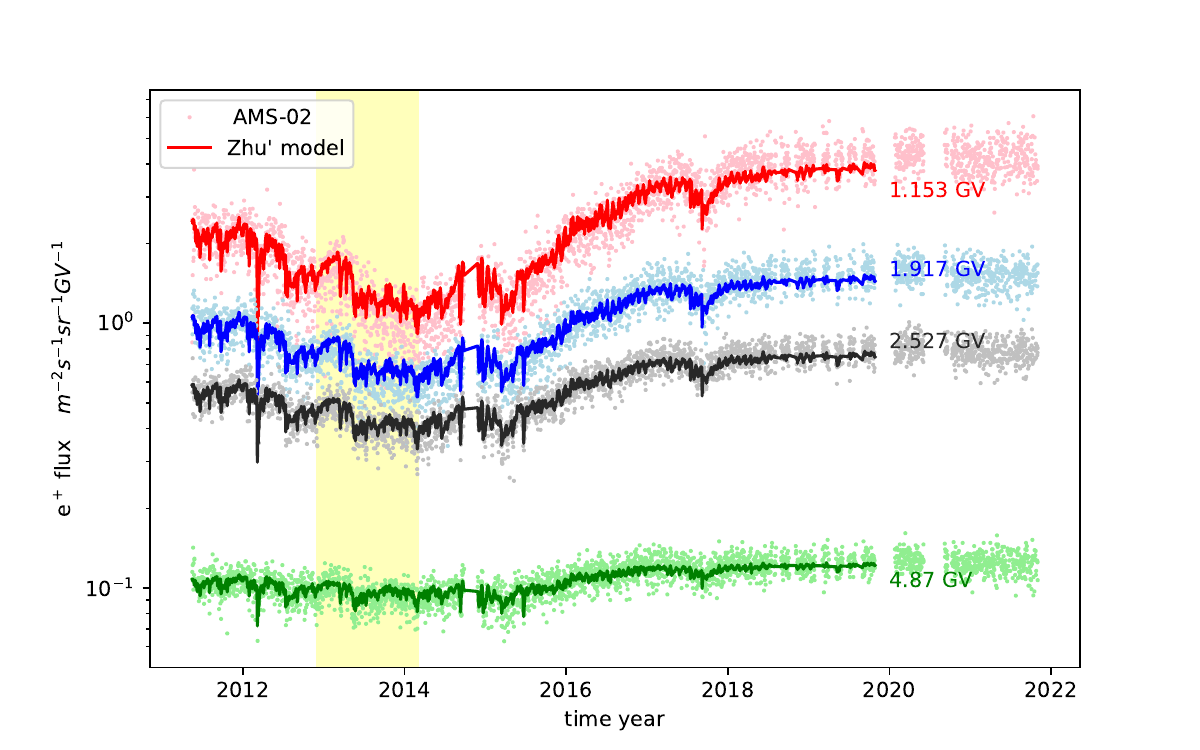}
    \caption{Same as Figure.~\ref{fig:posi_f} but for Zhu's model.}
    \label{fig:posi_f2}
\end{figure*}

In previous work, we demonstrated that the time-dependent fluxes of protons and Helium can be well-fitted using the same solar modulation parameters~\citep{zhu2024,zhu2024b}. Here, we calculated the positron fluxes with the LIS obtained above using the solar modulation parameters derived from ~\cite{zhu2024b}. The $\chi^2$ values for the two modified FFA models are shown in Figure~\ref{fig:chi2_posi}. The mean $\chi^2$ values are 14.38 for Zhu’s model and 13.37 for Long’s model, with 12 data bins. The higher $\chi^2$ values compared to direct positron data fitting suggest either the LIS requires refinement or the modified FFA models do not fully account for subtle differences in proton and positron modulation. Figure~\ref{fig:posi_f} and~\ref{fig:posi_f2} compares the positron fluxes predicted by Long's and Zhu's models with the AMS-02 data. Both models provide similar fitting results and agree well with the observed data.

 \subsection{Prediction of the Time-Dependent Fluxes of Antiprotons }

Building on these findings, we now turn to predicting the time-dependent fluxes of antiprotons. As demonstrated earlier, the time-dependent fluxes of protons and positrons can be fitted using the same solar modulation parameters when the appropriate LIS is employed. This consistency holds for both daily and monthly temporal resolutions. Given the analogous relationship between electrons and antiprotons, we can derive the daily antiproton fluxes from the daily electron fluxes.

\begin{figure*}
    \centering
    \includegraphics[scale=0.7]{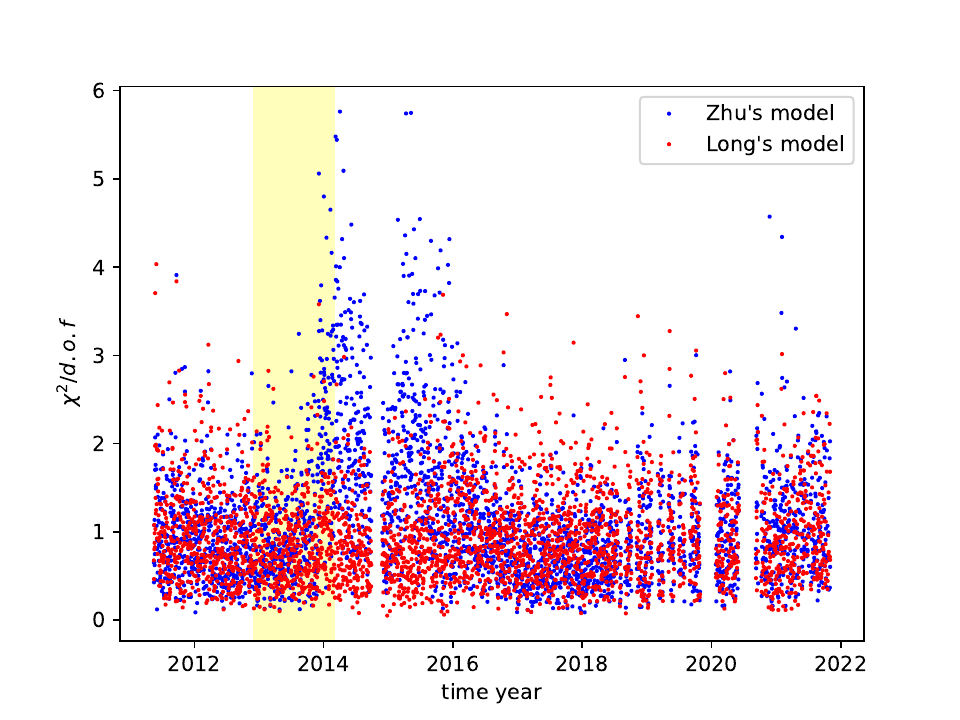}
    \caption{$\chi^2/d.o.f$ values from fitting to the daily electron fluxes for Zhu and Long's model. The yellow shaded band stands for the  heliospheric magnetic field reversal period within which the polarity is uncertain\citep{2015ApJ...798..114S}.}
    \label{fig:chi2e}
\end{figure*}

First, we fitted the daily electron fluxes using the LIS obtained earlier and the two modified FFA models. The $\chi^2$/d.o.f values are shown in Figure~\ref{fig:chi2e}. The mean $\chi^2$/d.o.f values are 1.143 for Zhu's model and 0.957 for Long's model. The results indicate that Long's model performs better, particularly one year after the heliospheric magnetic field reversal period, likely due to its ability to capture rigidity-dependent effects more accurately during this phase. Interestingly, this difference is not observed in the monthly electron fluxes. Figure~\ref{fig:chi2_mn} presents the $\chi^2$ values for Zhu's and Long's models using monthly electron fluxes, compared to the FFA model. Both models exhibit very similar fitting results, with mean $\chi^2$ values of 0.941 for Zhu's model and 0.946 for Long's model.

\begin{figure*}
    \centering
    \includegraphics[scale=0.7]{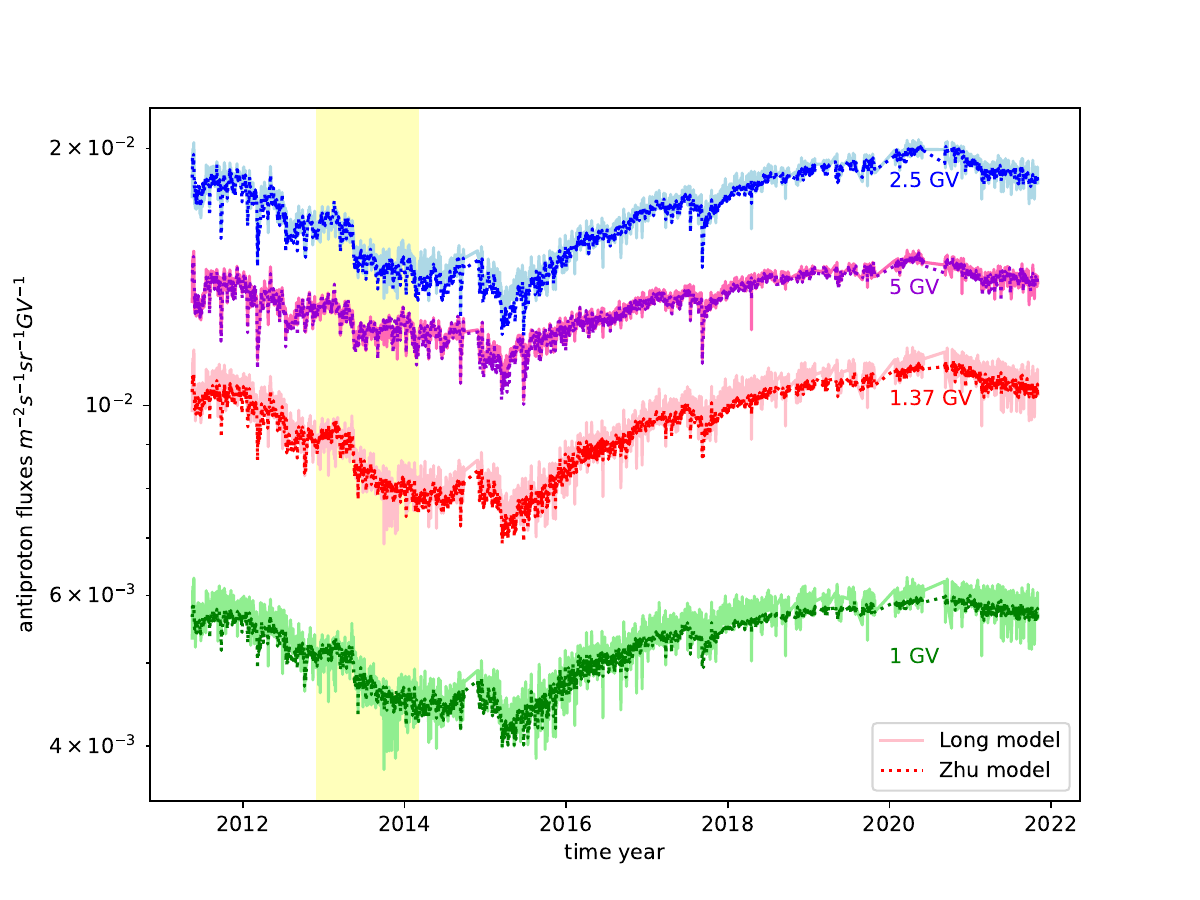}
    \caption{Model prediction  antiproton fluxes of Zhu (dash lines) and Long's model (lines) at rigidities = 2.5 GV, 5 GV, 1.37 GV and 1 GV (from top to bottom). The yellow shaded band stands for the  heliospheric magnetic field reversal period within which the polarity is uncertain\citep{2015ApJ...798..114S}.}
    \label{fig:pbar}
\end{figure*}

Next, we calculated the daily antiproton fluxes using the same solar modulation parameters derived from the electron data. The results for Zhu's and Long's models are shown in Figure~\ref{fig:pbar}. Both models produce similar antiproton flux series, though Long's model shows greater daily fluctuations. Long's model yields slightly higher mean fluxes than Zhu's model at low rigidities. Below approximately 3 GV, the antiproton fluxes increase with rigidity, reflecting the interplay between rigidity-dependent modulation effects and the energy spectrum of antiprotons. Above 3 GV, the antiproton fluxes decrease with rigidity, and the amplitude of flux variations diminishes as rigidity increases.  We have also forecasted the monthly fluxes of antiprotons using the monthly electron fluxes, and the results are very similar to the daily forecasts. 

\begin{figure*}
    \centering
    \includegraphics[scale=0.6]{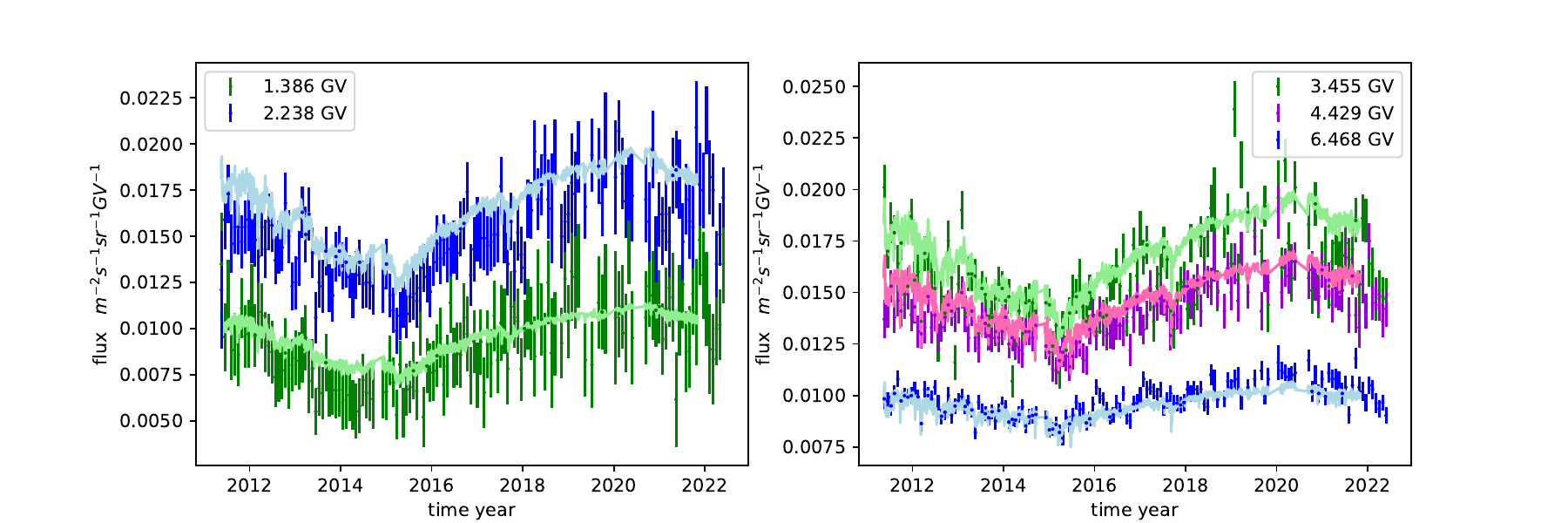}
    \caption{Zhu's model predictions  for antiproton fluxes (colored lines) are compared with  AMS-02 measurements (data points with error bars) \citep{PhysRevLett.134.051002}.}
    \label{fig:ap}
\end{figure*}

\begin{figure*}
    \centering
    \includegraphics[scale=0.6]{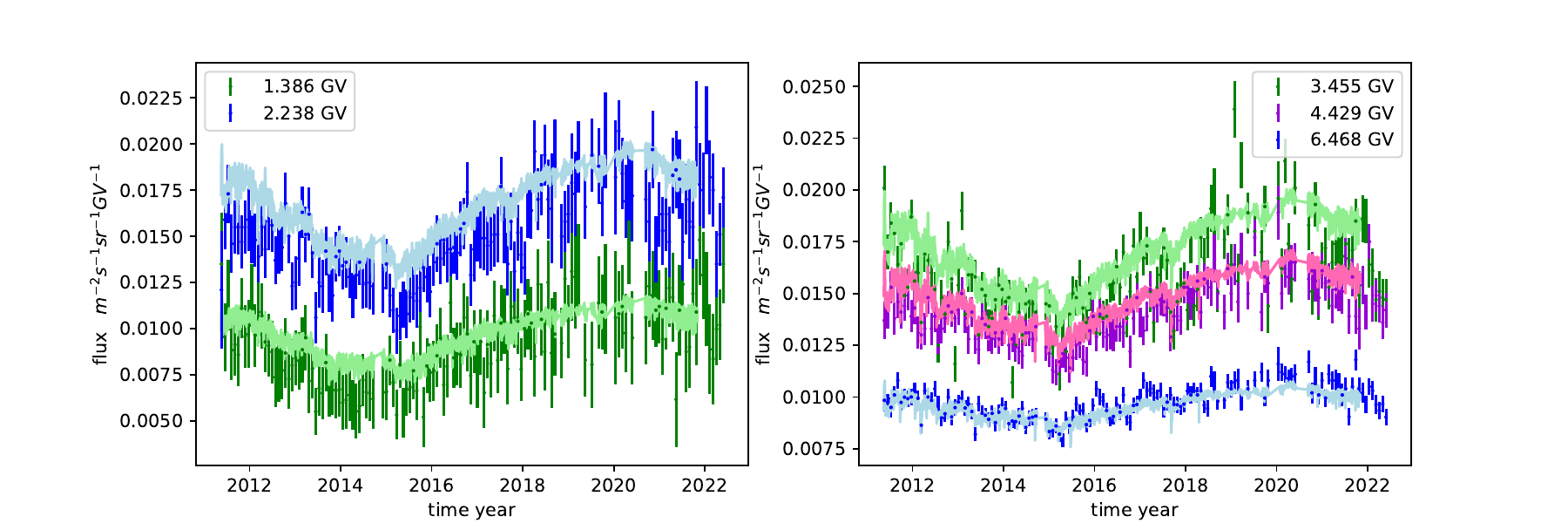}
    \caption{Long's model predictions for antiproton fluxes (colored lines) are compared with  AMS-02 measurements (data points with error bars) \citep{PhysRevLett.134.051002}.}
    \label{fig:ap2}
\end{figure*}

\zcr{During the review process of this work, we noticed that the AMS Collaboration's published antiproton flux measurements \citep{PhysRevLett.134.051002}, which coincided with our analysis timeline. Subsequently, we conducted systematic comparisons between our modified FFA model predictions and these benchmark experimental results in Figures \ref{fig:ap} and \ref{fig:ap2}. Remarkably, both modified FFA models demonstrate complete consistency with the latest AMS-02 data within the 1 $\sigma$ confidence interval. This empirical validation conclusively demonstrates the validity of our methodological framework in resolving solar modulation  of different GCRs.}

\begin{figure*}
    \centering
    \includegraphics[scale=0.9]{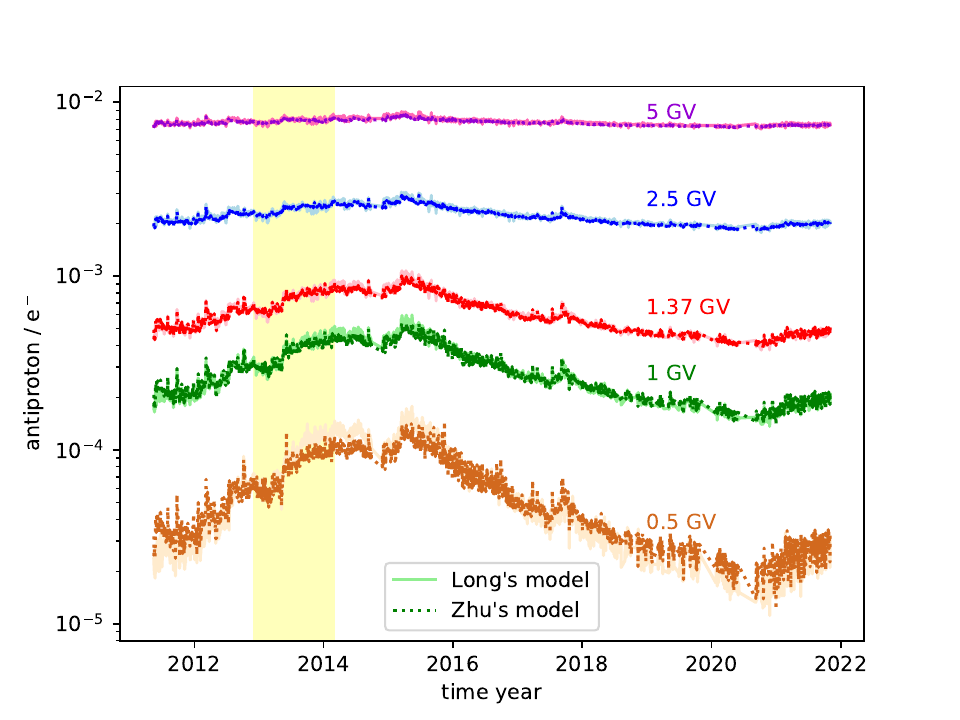}
    \caption{Model prediction  antiprotons  to electrons fluxes ratio of Zhu (dash lines) and Long's model (lines) at rigidities =  5 GV, 2.5 GV, 1.37 GV  1 GV  and 0.5 GV (from top to bottom). The yellow shaded band stands for the  heliospheric magnetic field reversal period within which the polarity is uncertain\citep{2015ApJ...798..114S}.}
    \label{fig:ratio}
\end{figure*}

Figure~\ref{fig:ratio} illustrates the model-predicted antiproton-to-electron flux ratio. The ratio increases monotonically with rigidity. At lower rigidities, the ratio exhibits similar evolutionary trends: it increased before 2015, decreased from 2015 to 2021 during solar maximum, and began to rise again after 2021. As rigidity increases, the relative fluctuation amplitude decreases. The evolution of the ratio is primarily influenced by two factors: the different shapes of the LIS and the differing kinetic-energy-to-rigidity conversions for electrons and antiprotons. According to the Equation~\ref{force_filed}, the ratio of antiprotons to electrons is given by:
 
\begin{equation}\label{eq:ratio}
\begin{aligned}
\frac{J_ {\bar p} ^{TOA}(R^{TOA})}{J_{e}^{TOA}(R^{TOA})} =& \frac{J_{\bar p}^{TOA}(E_{\bar p}^{TOA}) (\frac{Z\beta}{A})_{\bar p}}{J_{e}^{TOA}(E_{e}^{TOA}) (\frac{Z\beta}{A})_{e}}\\
 =& \frac{(\frac{Z\beta}{A})_{\bar p}}{(\frac{Z\beta}{A})_{e}}  \frac{J_{\bar p}^{LIS}(E_{\bar p}^{LIS})}{J_{e}^{LIS}(E_{e}^{LIS})} \left ( \frac{R_{e}^{LIS}}{R_{\bar p}^{LIS}} \right ) ^2.
 \end{aligned}
\end{equation}
where $E^{TOA} = E^{LIS } - \frac{Ze}{A}\phi(R)$,  $E = \sqrt{R^2(\frac{Ze}{A})^2+m_0^2}-m_0$, \( m_0 = 0.938 \) GeV for antiprotons and \( m_0 = 0.000511 \) GeV for electrons. This equation describes the relationship between the TOA energy and the LIS energy, accounting for solar modulation effects. With the same solar modulation parameters, the second and third terms in the Equation~\ref{eq:ratio} introduce time-dependent effects.  One can refer to \cite{zhu2024} for more details.


 \begin{figure*}
    \centering
    \includegraphics[scale=0.7]{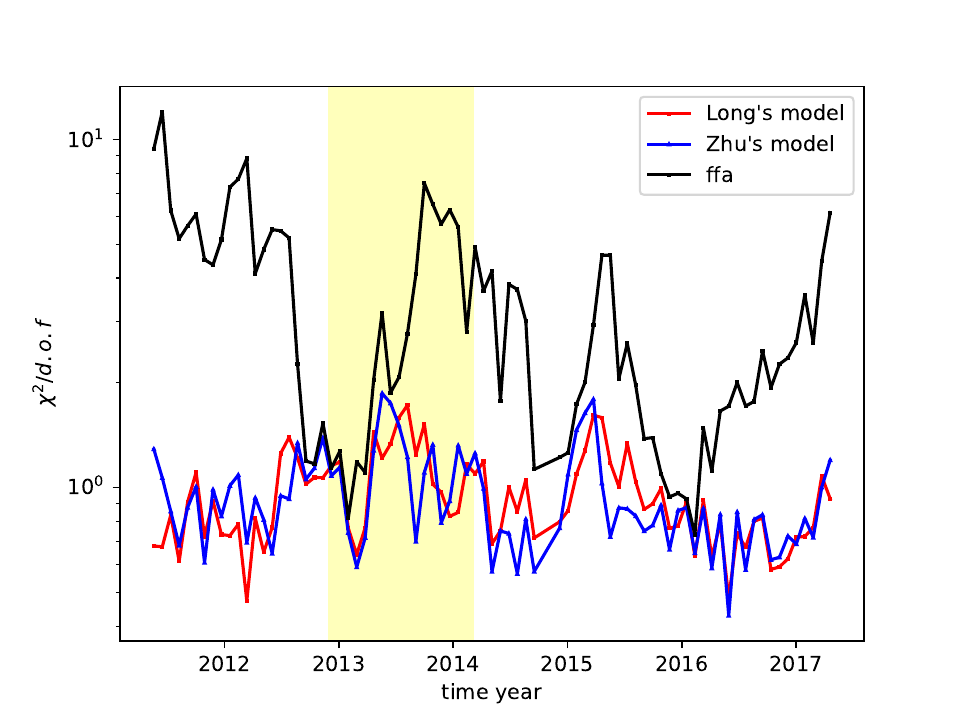}
    \caption{$\chi^2/d.of.$ values for Zhu (blue) and Long's (red) model compared to the FFA (black) using monthly proton and positron fluxes.  The yellow shaded band stands for the  heliospheric magnetic field reversal period within which the polarity is uncertain\citep{2015ApJ...798..114S}.}
    \label{fig:chi2e_mn}
\end{figure*}

\section{discussion and conclusion}

The precise daily measurements of cosmic ray spectra by AMS-02 provide a unique opportunity to enhance our understanding of solar modulation. In this study, we model the daily fluxes of protons, positrons, and electrons from AMS-02 using modified FFA models and predict the time-dependent fluxes of antiprotons during the AMS-02 era.

First, we establish that protons and positrons exhibit identical solar modulation parameters. By simultaneously fitting the time-dependent fluxes of protons and positrons using two modified FFA models, we achieve a $\chi^2/d.o.f$ value of approximately 0.75. Since protons and antiprotons have the same mass but opposite charges, and positrons and electrons also have the same mass but opposite charges, we assume that antiprotons and electrons share the same solar modulation parameters. Based on this assumption, we extract the solar modulation parameters from the time-dependent electron fluxes and use them to calculate the antiproton fluxes with the two modified FFA models. Our results are presented for the rigidity range of 1 to 5 GV, with additional data at more rigidities accessible on our homepage\footnote{\url{https://github.com/dkkyjy/the-daily-fluxes-of-antiprotons}}.

We also analyze the antiproton-to-electron flux ratio in the rigidity range of 0.5 to 5 GV. The results reveal distinct temporal trends: an initial increase before 2015, followed by a decrease from 2015 to 2021, and a subsequent rise after 2021. As rigidity increases, the relative fluctuation amplitude of the ratio decreases. These long-term trends are primarily influenced by variations in the LIS shapes and the rest-mass energies of antiprotons and electrons. As indicated by the Equation~\ref{eq:ratio}, the second and third terms on the right-hand side, combined with the same solar modulation parameters, account for these long-term behaviors.

This study is based on the hypothesis that GCRs with the same charge sign exhibit similar solar modulation parameters due to similar diffusion and energy loss processes in the heliosphere. The time-dependent fluxes of protons, helium, and positrons validate this hypothesis for positively charged GCRs. 
\zcr{The latest AMS-02 time-dependent antiproton flux measurements  provide a critical test for negatively charged GCRs and this finding will profoundly enhance our understanding of the origin and propagation of GCRs in the galaxy. This finding will also be useful for forecasting the time-dependent fluxes of other GCR species, such as Ne, Mg, Si and so on.} 

\begin{acknowledgments}
Thanks for Fan Yi-Zhong, Yuan Qiang  and Shu Xin-Wen for very helpful discussions. This work is supported by the National Natural Science Foundation of China  (No. 12203103).  D.K.K. is supported by the National Key Research and Development Program of China (No. 2022YFF0503304). Z.C.R is also supported by the Doctoral research start-up funding of Anhui Normal University. We acknowledge the use of  data from the {AMS Publications} \footnote{\url{https://ams02.space/publications/}} and the {Cosmic-Ray Database} \citep{DiFelice:2017Hm}~\footnote{\url{https://tools.ssdc.asi.it/CosmicRays/}}. 
\end{acknowledgments}

\clearpage 

\bibliographystyle{aasjournal}
\bibliography{refs}

\end{document}